\documentclass[11pt]{article}

\usepackage{latexsym,bm,graphicx,color,xcolor,nicefrac,titletoc,enumerate,amsmath,amssymb,xfrac,xcolor,physics,cite,setspace}

\usepackage{booktabs}        
\usepackage{threeparttable}  

\usepackage{mlmodern}
\usepackage[T1]{fontenc}

\usepackage[nottoc]{tocbibind} 

\usepackage{hyperref}
\hypersetup{linktocpage=true,colorlinks=true,linkcolor=blue,citecolor=blue,urlcolor=blue}

\usepackage[skip=3pt plus1pt, indent=20pt]{parskip}

\usepackage[bottom]{footmisc}

\usepackage{geometry}
\usepackage{titlesec}
\titleformat{\section}{\large\bfseries\sffamily}{\thesection}{0.5em}{}
\titleformat{\subsection}{\normalfont\bfseries\sffamily}{\thesubsection}{0.5em}{}
\titleformat{\subsubsection}{\normalsize\itshape\sffamily}{\thesubsubsection}{0.5em}{}
\titleformat*{\paragraph}{\normalsize\bfseries\sffamily}

\numberwithin{equation}{section}

\def\a{\alpha}

\def\g{\gamma}

\def\f{\phi}
\def\vf{\varphi}

\def\l{\lambda}
\def\L{\Lambda}
\def\m{\mu}
\def\n{\nu}
\def\r{\rho}
\def\s{\sigma}

\def\th{\theta}

\def\x{\xi}
\def\z{\zeta}

\def\pd{\partial}

\def\pr{\prime}

\def\nn{\nonumber}

\newcommand{\sq}{\sqrt}
\newcommand{\sqdet}{\sq{-g}}

\newcommand{\cL}{\mathcal{L}}

\newcommand{\cN}{\mathcal{N}}
\newcommand{\cR}{\mathcal{R}}

\newcommand{\Lg}{\cL_\text{g}}
\newcommand{\Lm}{\cL_\text{m}}
\newcommand{\Ired}{I_{\text{red}}}
\newcommand{\Lred}{L_{\text{red}}}
\newcommand{\Ilog}{I_{\text{log}}}

\newcommand{\mail}[1]{\href{mailto:#1}{{\tt #1}}}

\begin{document}

\begin{titlepage}       
	
	
	\begin{center}
		{\Large \bf \sffamily 3d Lovelock gravity and the holographic c-theorem:\\ Proof at all orders and resummation}
	\end{center}
	
	\begin{center}
		\vspace{10pt}
		
		{{\bf \sffamily G{\"o}khan Alka\c{c},}${}^{a}\,${\bf \sffamily  Luis Guajardo}${}^{b}\,$ {\bf \sffamily and Hikmet \"{O}z\c{s}ahin}${}^{c}$}
		\\[4mm]
		
		{\small 
			{\it ${}^a$Department of Aerospace Engineering, Faculty of Engineering,\\ At{\i}l{\i}m University, 06836 Ankara, T\"{u}rkiye}\\[2mm]
			
			{\it ${}^b$Instituto de Matem\'atica, F\'isica y Estad\'istica, Facultad de Ingenier\'ia y Negocios, \\Universidad de Las Am\'ericas, Sede Concepci\'on, Avenida Jorge Alessandri Rodr\'iguez 1160, 4090940, Chile}\\[2mm]
			
			{\it ${}^c$Department of Physics, Faculty of Arts and Sciences,\\ Sinop University, 57000 Sinop, T\"{u}rkiye}\\[2mm]
			
			{\it E-mail:} {\mail{alkac@mail.com}, \mail{luis.guajardo.r@gmail.com}, \mail{h.ozsahin@sinop.edu.tr}}
		}
		\vspace{2mm}
	\end{center}
	

		\noindent We prove that 3d Lovelock gravity, the Horndeski theory obtained from the regularized Lovelock invariants in three dimensions, admits a holographic c-theorem at all orders and for arbitrary values of the couplings. In the minisuperspace of domain wall solutions, the equation of the Horndeski scalar is a total derivative at every order, which locks the derivative of the scalar to that of the warp factor along the flows we consider and allows the null energy condition to be integrated into a monotonic c-function. The linear scalar profile of the resulting vacuum breaks the special conformal transformations, so the dual theory is scale-invariant but not conformal, and the central charge is defined as the coefficient of the logarithmic divergence of the on-shell action on the sphere. We compute it exactly for the Gauss-Bonnet case and asymptotically for the full tower, and find that it is linear in the couplings and matches the UV value of the c-function. Since all results depend on the couplings only through two generating functions, we give the resummed Lagrangians, c-functions and central charges in closed form for the choices of couplings that lead to regular black holes, and classify the resummed theories by the positivity of the effective Newton constant on the vacuum and the uniqueness of the vacuum with a non-trivial scalar.

	\tableofcontents
\end{titlepage}


\section{Introduction}\label{sec:intro}
General relativity (GR) is a remarkably successful theory, but it is also expected to be an effective description valid at low energies. Higher-curvature corrections to the Einstein-Hilbert action arise generically, for instance from string theory \cite{Zwiebach:1985uq, Gross:1986iv, Gross:1986mw, Metsaev:1987zx, Boulware:1985wk}, and studying their consequences is the standard way to test which properties of GR are robust and which are accidents of the two-derivative truncation. In this context, the choice of the correction terms matters: a generic higher-curvature term produces higher-order field equations, and with them additional degrees of freedom that are typically ghost-like \cite{Stelle:1976gc, Stelle:1977ry, Woodard:2015zca}.

Lovelock gravity \cite{Lovelock:1971yv,Lovelock:1972vz} is the most natural modification of GR in this respect. Its Lagrangian is a sum of the densities
\begin{equation}
	\cL_n =\frac{1}{2^n}\delta^{\mu_1\nu_1\cdots\mu_n\nu_n}_{\rho_1\sigma_1\cdots\rho_n\sigma_n}R^{\rho_1\sigma_1}_{\ \ \ \ \mu_1\nu_1}\cdots R^{\rho_n\sigma_n}_{\ \ \ \ \mu_n\nu_n},\label{Ln}
\end{equation}
where $\delta^{\mu_1\cdots\mu_{2n}}_{\rho_1\cdots\rho_{2n}}$ is the generalized Kronecker delta, so that $\cL_1 = R$ and $\cL_2 = R^2 - 4R_{\m\n}^2 + R_{\m\n\r\s}^2$ is the Gauss-Bonnet invariant. It is the unique theory built from the metric alone whose field equations are of second order. It admits static, spherically symmetric black hole solutions in closed form and, around the vacuum continuously connected to GR, it propagates only a massless spin-2 graviton, which is unitary \cite{Sisman:2012rc} (see \cite{Padmanabhan:2013xyr} for a review of its many other properties). The price for these properties is that, for $d<5$, the only surviving member of the family is the Einstein-Hilbert term: the density \eqref{Ln} is antisymmetrized over $2n$ indices, so it vanishes identically for $d<2n$ and is a topological invariant for $d=2n$, contributing to the field equations only for $d>2n$.

Modified gravity theories are also an essential tool in the context of the AdS/CFT correspondence \cite{Maldacena:1997re,Witten:1998qj,Gubser:1998bc}. Since a gravitational theory with a negative cosmological constant defines a dual field theory, higher-curvature corrections allow one to explore a wider class of field theories than those dual to GR, and to test which holographic statements are universal. Lovelock and related theories have been particularly useful here. Notable examples are the studies of the shear viscosity bound and of causality constraints \cite{Brigante:2007nu,Brigante:2008gz,Buchel:2009sk,Camanho:2009vw,deBoer:2009pn,deBoer:2009gx,Camanho:2009hu}.

Three-dimensional (3d) gravity offers a simpler setup to study gravitational phenomena, and this is also true for the AdS$_3$/CFT$_2$ correspondence, thanks to the infinite number of symmetries of two-dimensional (2d) conformal field theories (CFTs); see \cite{Kraus:2006wn} for a review. In 3d, the closest one can get to Lovelock gravity with a theory built from the metric alone is New Massive Gravity (NMG) \cite{Bergshoeff2009,Bergshoeff:2009aq}, with the Lagrangian $\cL = -R + \frac{1}{m^2}\left[R_{\m\n} R^{\m\n} - \frac{3}{8} R^2\right]$. Despite having fourth-order field equations, it does not suffer from the problem mentioned above: the specific combination of curvature-squared terms removes the scalar mode, and the additional degrees of freedom are those of a unitary massive graviton of mass $m$. It has turned out to be a very useful toy model with many interesting applications.

From the holographic point of view, the analogy between NMG and Lovelock gravity was made precise by Sinha \cite{Sinha:2010ai}. The starting point is the holographic c-theorem of \cite{Freedman:1999gp}: along a renormalization group (RG) flow driven by matter fields obeying the null energy condition (NEC), one can construct a function of the radial coordinate that is monotonic, and whose value at the UV fixed point ($r\to\infty$) is proportional to the Weyl anomaly coefficient of the boundary theory \cite{Henningson:1998gx}, when the boundary is even-dimensional. This is the holographic version of Zamolodchikov's c-theorem for 2d quantum field theories, $c_{\rm UV} > c_{\rm IR}$ for any RG flow connecting two fixed points \cite{Zamolodchikov:1986gt}. A natural question is which theories of gravity admit a holographic c-theorem under the same assumptions. Sinha considered the most general four-derivative deformation of GR in 3d, $\cL = R + a R^2 + b R_{\m\n} R^{\m\n}$ with constants $a$ and $b$, and showed that NMG is the unique theory in this class that admits a holographic c-theorem. This was later extended to an arbitrary number of curvature corrections\footnote{In $d>4$, besides Lovelock gravity, quasi-topological gravity \cite{Myers:2010ru,Oliva:2010eb} also admits a holographic c-theorem. Its field equations are of higher order in general, but become second order for static, spherically symmetric metrics. Myers and Sinha used it to give a physical meaning to the UV value of the c-function when the boundary theory is odd-dimensional, completing the proof of the holographic c-theorem in arbitrary dimensions \cite{Myers:2010xs,Myers:2010tj}. We do not need this here: the boundary theory is two-dimensional, and the coefficient of the logarithmic divergence of the on-shell action is all that is required.} in \cite{Paulos:2010ke} (see \cite{Gullu:2010st, Alkac:2018whk} for Born-Infeld-type extensions). The 3d theories admitting a holographic c-theorem were classified in \cite{Bueno2022} (see \cite{Bergshoeff:2021tbz} for their construction from truncations of an infinite-dimensional extension of the AdS$_3$ algebra, which relates them to the Chern-Simons-like formulation of massive gravity), and the subclass with a unique AdS$_3$ vacuum was constructed at arbitrary order in the curvature in \cite{Chernicoff:2024lkj}, where it was shown that the ghost mode of these theories decouples around AdS$_3$ and that their central charge is set by an effective Planck length.

One may still object that fourth-order field equations are an undesirable feature, and that NMG and its higher-order extensions are not the true analogues of Lovelock gravity in 3d. Recently, it was shown that a well-defined limit of the Lovelock densities \eqref{Ln} to lower dimensions can be taken, through different regularization procedures \cite{Fernandes:2020nbq,Lu:2020iav,Kobayashi:2020wqy,Hennigar:2020lsl,Alkac:2022fuc,Bravo-Gaete:2022mnr}. The result is a Horndeski theory \cite{Horndeski:1974wa,Kobayashi:2019hrl}, i.e.\ a scalar-tensor theory with second-order field equations.\footnote{Along similar lines, vector-tensor theories with second-order field equations, of the generalized Proca type \cite{Heisenberg:2014rta}, have also been obtained and studied in \cite{Charmousis:2025jpx,Eichhorn:2025pgy,Alkac:2025zzi,Liu:2025dqg,Alkac:2025jhx,Lutfuoglu:2025qkt,Konoplya:2025bte,Charmousis:2026dbi}.} The scalar field is a remnant of the extra dimensions, and the resulting theories inherit many properties of their higher-dimensional parents; see \cite{Ma:2020ufk,Hennigar:2020fkv,Hennigar:2020drx,Alkac:2023mvr} for the study of their vacua, black hole solutions and thermodynamics. In this paper, we call the 3d theories obtained in this way 3d Lovelock gravity.

The regularization procedure was extended to arbitrary order in the Lovelock series in \cite{Colleaux:2020wfv}, where the resulting Horndeski Lagrangians were obtained in closed form, and more recently it was applied in 4d and 3d in \cite{Fernandes:2025fnz,Fernandes:2025eoc}. It was shown in \cite{Alkac:2022zda} that 3d Lovelock gravity admits a holographic c-theorem up to cubic order, which supports the view that these theories, rather than NMG, are the true 3d analogues of Lovelock gravity. The all-order Lagrangians opened the possibility, first explored in 4d quasi-topological gravity \cite{Bueno:2024dgm}, of resumming the infinite tower of corrections for particular choices of the coupling constants. The resummed theories have remarkable properties: the black hole and cosmological singularities are resolved \cite{Fernandes:2025fnz,Fernandes:2025eoc}, and the thermodynamics of the resulting regular black holes is well defined \cite{Cisterna:2025vxk}. Their microscopic entropy in 3d has been derived from a Cardy formula with the soliton as the ground state \cite{Ozsahin:2026eao}.

In this work, we fill an important gap in the study of these theories. We prove that 3d Lovelock gravity admits a holographic c-theorem at all orders, for arbitrary values of the coupling constants, and we verify the result by an independent computation of the central charge, as the coefficient of the logarithmic divergence of the Euclidean on-shell action on the sphere. Both computations are done in a minisuperspace approach, where the field equations follow from a reduced action for a symmetric ansatz. This is what makes the all-order proof possible: the reduced action is easy to compute for the whole tower, and the key equations turn out to be total derivatives. Along the way, we find that the scalar profile of the AdS$_3$ vacuum breaks the special conformal transformations, so that, as far as the symmetries of the vacuum are concerned, the dual field theory is scale-invariant but not conformal. The c-theorem and the computation of the central charge are unaffected, but the usual identification of the central charge with the Weyl anomaly coefficient of a CFT should be understood in this sense. We then apply the result to the resummed theories. Since the c-function and the central charge depend on the couplings only through a generating function, they can be given in closed form for particular choices of the $c_n$. We do this for the choices used to construct regular black holes in 3d \cite{Fernandes:2025eoc}. Positivity of the central charge, which in the bulk is the positivity of the effective Newton constant on the vacuum, constrains the resummed theories.

The paper is organized as follows. In Section~\ref{sec:toy}, we introduce the method in a simpler Horndeski model, where the warp factor does not run along the flow, compute both the c-function and the central charge from the on-shell action, and discuss the symmetries of the vacuum. In Section~\ref{sec:proof}, we give the general proof for the full tower of regularized Lovelock invariants: we find the vacuum, construct the c-function and check its UV value against the on-shell action. In Section~\ref{sec:resum}, we resum the tower for the choices of couplings studied in 3d \cite{Fernandes:2025eoc}, give the resummed c-functions and central charges, and discuss the positivity of the central charge and the structure of the vacua. Section~\ref{sec:discussion} contains a discussion of the results and possible future directions.

\section{A simpler model}\label{sec:toy}
In this section, to introduce the methods we will use, we study a simpler Horndeski model. It is defined by the action
\begin{equation}\label{actg}
	I = \int \dd^3 x \sqdet \left[R - 2\L_0 - \frac{1}{2} \left(\a g^{\m\n} - \g G^{\m\n}\right) \pd_\m \f \pd_\n \f \right],
\end{equation}
where $\a$ and $\g$ are coupling constants and $\L_0$ is the bare cosmological constant. In 3d, this model was first studied in \cite{Bravo-Gaete:2014haa} in the context of black hole thermodynamics. The authors showed that the Ba\~nados-Teitelboim-Zanelli black hole \cite{Banados:1992wn} is a solution with a non-trivial scalar profile, which modifies the thermodynamic properties compared to pure GR with a cosmological constant (see \cite{Anabalon:2013oea} for the same model in 4d). The c-theorem in this model was studied in \cite{Li:2018kqp}. Here, however, we adopt the minisuperspace approach introduced in \cite{Li:2017txk}, since it can handle more complicated models and prepares us for 3d Lovelock gravity.

For a general gravitational theory with Lagrangian $\Lg$, we need to add matter fields with a Lagrangian $\Lm$ obeying the NEC, and then check whether a monotonic c-function can be constructed. Varying the action
\begin{equation}\label{actgen}
	I = \int \dd^3 x \sqdet \left(\Lg + \Lm \right),
\end{equation}
with respect to the metric gives field equations of the general form
\begin{equation}\label{eqns}
	\Psi_{\m\n} = T_{\m\n},
\end{equation}
where $\Psi_{\m\n}$ comes from $\Lg$ and $T_{\m\n}$ is the energy-momentum tensor of $\Lm$.

To construct the c-function, we consider the domain wall ansatz
\begin{equation}\label{dom}
	\dd{s}^2 = e^{2A(r)} (-\dd{t}^2 + \dd{x}^2) + \dd{r}^2.
\end{equation}
The NEC, $T_{\m\n} \z^\m \z^\n \geq 0$ for any null vector $\z^\m$, can be imposed with the choice $\z^\m = (\z^t,\z^x,\z^r) = v\,(e^{-A(r)},0,1)$, where $v$ is an arbitrary function. The NEC then reads
\begin{equation}\label{NEC1}
	T^r_{\ r} - T^t_{\ t} \geq 0,
\end{equation}
which implies
\begin{equation}\label{NEC2}
	\Psi^r_{\ r} - \Psi^t_{\ t} \geq 0.
\end{equation}

For GR with a negative cosmological constant ($\L_0 = -1/\ell^2$), one has
\begin{equation}\label{GReqs}
	\Lg = R - 2 \L_0, \qquad \Psi_{\m\n} = G_{\m\n} + \L_0 g_{\m\n}, \qquad - A^{\pr \pr} \geq 0,
\end{equation}
where $^\pr$ denotes the derivative with respect to the radial coordinate $r$. By inspection, one can define the monotonically increasing c-function
\begin{equation}\label{cGRfunc}
	c(r) = \frac{1}{\ell_p A^\pr}.
\end{equation}
At the UV boundary ($r \to \infty$), the matter fields should vanish and one has the AdS$_3$ vacuum solution\footnote{Defining the new radial variable $z= \ell e^{-r/\ell}$ in the domain wall ansatz \eqref{dom} with $A=r/\ell$, one obtains the metric of AdS$_3$ in the Poincar\'e patch, $\dd{s}^2 = \frac{\ell^2}{z^2} (-\dd{t}^2 + \dd{x}^2 + \dd{z}^2)$.}
\begin{equation}\label{GRvac}
	A = \frac{r}{\ell}, \qquad \L_0 = -\frac{1}{\ell^2}.
\end{equation}

The value of the c-function at the boundary is the central charge of the dual CFT, which in this case is\footnote{We use a normalization different from that of Brown and Henneaux \cite{Brown:1986nw}, where $c = \frac{3 \ell}{2 G_3}$, so that $\ell_p = 2G_3/3$.}
\begin{equation}\label{cGR}
	c_\infty = \frac{\ell}{\ell_p}.
\end{equation}
In two dimensions, the central charge of the Virasoro algebra is also the coefficient of the Weyl anomaly, $\langle T^\m_{\ \m} \rangle = -\frac{c}{24\pi} R$, so that on a curved boundary the partition function of the CFT acquires a logarithmic dependence on the UV cut-off with a coefficient fixed by $c$. For a sphere of radius $a$, $\log Z = \frac{c}{3}\log a + \dots$. Holographically, this logarithm appears as a divergence of the on-shell gravitational action \cite{Henningson:1998gx}. This provides an independent check of \eqref{cGR}, which we perform later.

Before doing that, we explain how this analysis can be carried out in the minisuperspace of the relevant field configurations. This is very useful, because computing the covariant field equations \eqref{eqns} for a general model can be quite involved. For the metric we take the slightly more general form
\begin{equation}\label{domgen}
	\dd{s}^2 = e^{2A(r)} (-\dd{t}^2 + \dd{x}^2) + e^{2B(r)} \dd{r}^2,
\end{equation}
and a radial profile for the Horndeski scalar, $\f = \f(r)$. For the matter sector we introduce a free scalar,
\begin{equation}\label{Lm}
	\Lm= - \frac{1}{2} (\pd \chi)^2,
\end{equation}
so that the NEC \eqref{NEC1} becomes the positivity of the radial kinetic energy:
\begin{equation}\label{Fdef}
	T^r_{\ r} - T^t_{\ t} = \frac{1}{2} \chi'^2 \equiv F \geq 0.
\end{equation}
A free scalar has no potential, so the flow it drives does not end at an IR fixed point. This is not needed: as in \cite{Freedman:1999gp,Sinha:2010ai}, the holographic c-theorem is the statement that the c-function is monotonic along any flow driven by matter obeying the NEC. Inserting these field configurations into the action \eqref{actgen}, one finds a reduced action $\Ired = \int \dd{r} \Lred[\Phi_A]$ for the functions $\Phi_A = \{A,B,\f,\chi\}$. This reduced action is easy to compute, and it is enough to find the vacuum solution of the theory and to study the behaviour of the functions $A$ and $\f$ when the NEC is imposed.

Two comments are in order about this procedure. First, restricting the action to the ansatz \eqref{domgen} before varying is legitimate because this ansatz is the most general field configuration invariant under the symmetries of the domain wall (translations in $t$ and $x$ and the boost in the $t$-$x$ plane). By the principle of symmetric criticality \cite{Palais1979,Fels2002}, the Euler-Lagrange equations of the reduced action coincide with the full field equations evaluated on the ansatz. Second, although the gauge $g_{rr}=1$ could be imposed directly in \eqref{dom} when working with the covariant field equations \eqref{eqns}, where the constraint is just the $rr$ component, this is not possible when the equations are derived from a reduced action. The function $B$ must be kept and varied, and $B=0$ can only be imposed afterwards. The reason is that $e^{B}$ plays the role of a lapse: it appears in $\Lred$ without derivatives, and its equation of motion is the Hamiltonian constraint. This equation is not implied by the others, which only guarantee that the constraint is constant along $r$. Setting $B=0$ in the action would lose it, and with it the relation between the bare cosmological constant and the AdS radius. Once all the equations are obtained, $B=0$ is simply a choice of the radial coordinate.

Let us apply this procedure to the model \eqref{actg}. After deriving the Euler-Lagrange equations $\fdv{\Ired}{\Phi_A}= 0$, we set $B =0$ and find
\begin{align}
	0 &= 4\Lambda_0 + {\chi'}^2+ 2\gamma \phi' \phi'' A'+ (4 + \gamma {\phi'}^2) {A'}^2 + \phi'^2 \left(\alpha + \gamma A''\right)  +4A'',\label{Aeqn}\\
	0 &= 4\Lambda_0 - \chi'^2	- \alpha{ \phi'}^2 + (4 + 3\gamma {\phi'}^2)   A'^2,\label{Beqn}\\
	0 &= \alpha \phi''-2\gamma \phi' {A'}^3  - \gamma \phi'' {A'}^2  + 2\phi' A'  (\alpha - \gamma A''),\label{phieqn}\\
	0 &=  2\chi' A'  + \chi''.\label{chieqn}
\end{align}

The vacuum AdS$_3$ solution with effective cosmological constant $\L = -\frac{1}{\ell^2}$ is found by setting $A=r/\ell$ and $\chi = 0$. This gives a relation between the coupling constants and a linear profile for the scalar field,
\begin{equation}\label{toyvac}
	\g = \a \ell^2, \qquad \phi = \sqrt{-\frac{2(1+\L_0 \ell^2)}{\a \ell^2}} \,r.
\end{equation}
Note that the scalar profile is linear in $r$. This has a consequence for the dual field theory. Since the action \eqref{actg} depends on $\f$ only through its derivatives, a Killing vector $\x$ of AdS$_3$ is a symmetry of the vacuum if it shifts $\f$ by a constant, i.e.\ if $\x^\m \pd_\m \f \propto \x^r$ is constant. Of the six Killing vectors of AdS$_3$, only four satisfy this condition: the three generators of the 2d Poincar\'e group, which have $\x^r = 0$, and the dilatation $-\ell\pd_r + t\pd_t + x\pd_x$. The two special conformal generators have an $r$-component that depends on $t$ and $x$, and are broken by the scalar. The dual field theory is therefore scale-invariant but not conformally invariant.  

For this model, the scale-invariant vacuum and its holographic two-point functions were studied in \cite{Li:2018kqp,Li:2018rgn} (see \cite{Chawla:2025fpn} for a general discussion of bulk geometries realizing scale without conformal invariance and a review of the literature). As discussed in \cite{Chawla:2025fpn}, in Einstein gravity with minimally coupled matter obeying the NEC, an AdS metric cannot be supported by matter that breaks the special conformal transformations, since the matter stress tensor would have to be proportional to the metric on the vacuum. However, the non-minimal coupling of the Horndeski scalar evades this, as the scalar terms in \eqref{actg} vanish identically on the vacuum. We also note that in \cite{Li:2018rgn} the two-point functions of the trace mode were found to vanish in $d=2$, so that the breaking of conformal invariance is not visible in the two-point functions of the stress tensor, consistently with the enhancement from scale to conformal invariance expected in two dimensions \cite{Polchinski:1987dy,Nakayama:2013is}.

The absence of full conformal invariance does not affect the c-theorem, which relies only on the NEC, nor the computation of the logarithmic divergence  of the on-shell action on S$^2$, which is well defined for any scale-invariant theory and reduces to the Weyl anomaly when the theory is conformal. In what follows, we define the central charge $c_\infty$ as the coefficient of this divergence, normalized so that it agrees with the Virasoro central charge in GR, and we keep the standard holographic terminology with this understanding.

To study the NEC, we eliminate the bare cosmological constant from Eqs.~(\ref{Aeqn}, \ref{Beqn}) and arrive at the inequality
\begin{equation}\label{modineq}
	F =  - \frac{1}{4} \phi'^2 (2\alpha - 2\gamma A'^2 + \gamma A'')- \frac{1}{2}\gamma  \phi' \phi''A' - A'' \geq 0.
\end{equation}
Remarkably, Eq.~\eqref{phieqn}, which follows from the variation with respect to $\f$, is a total derivative and integrates to
\begin{equation}\label{firstint}
	e^{2A} (\alpha - \gamma {A'}^2) \phi' = \lambda, \qquad \lambda = \text{constant}.
\end{equation}
Eq.~\eqref{phieqn} is the conservation of the current $J^\m = -(\a g^{\m\n} - \g G^{\m\n})\pd_\n\f$ of the shift symmetry $\f \to \f + \text{const}$, and $\l = -\sqrt{-g}\,J^r$ is the corresponding conserved charge. It is an independent parameter of the solutions, but a non-zero value is singular: the invariant $J_\m J^\m = \l^2 e^{-4A}$ diverges at the Poincar\'e horizon $r \to -\infty$ of the vacuum, where the warp factor vanishes although the geometry is regular. We therefore restrict to the sector $\l = 0$. Note that the theory also admits the GR vacuum $\f' = 0$ with $\L_0 = -1/\ell^2$ for any $\a$ and $\g$. Near this vacuum $\a - \g A'^2 \neq 0$, and $\l = 0$ gives $\f' = 0$ along the flow, so that one recovers the GR c-function \eqref{cGRfunc}. 

Around the vacuum \eqref{toyvac}, it follows from \eqref{firstint} that, wherever $\phi'\neq 0$,
\begin{equation}\label{Aexact}
	{A'}^2 = \frac{\a}{\g} = \frac{1}{\ell^2},
\end{equation}
i.e.\ the warp factor does not run along the flow, and the general solution for $A$ is just the vacuum solution. All the running is carried by $\phi'$ and $\chi'$. Using this in the inequality \eqref{modineq}, we obtain
\begin{equation}\label{Ftoy}
	F = -\frac{1}{2} \a \ell\, \f' \f'' \geq 0.
\end{equation}
Now consider the following proposal for the c-function,
\begin{equation}\label{toycfunc}
	c(r) = \frac{1}{\ell_p A'} \left[1-\frac{1}{4} \a \ell^2 {\f'}^2\right],
\end{equation}
which is monotonic since
\begin{equation}\label{toycprime}
	c'(r) = \frac{\ell^2}{\ell_p} F \geq 0.
\end{equation}
The central charge is obtained by evaluating the c-function \eqref{toycfunc} at infinity,
\begin{equation}\label{toyc}
	c_\infty = \frac{\ell(3 + \L_0 \ell^2)}{2 \ell_p}.
\end{equation}
In the GR limit, $\L_0 \ell^2 \to -1$ [see Eq.~\eqref{toyvac}], one recovers the result \eqref{cGR}. This fixes the constant term in the c-function \eqref{toycfunc}. Since $c'(r)$ is fixed by \eqref{toycprime}, \eqref{toycfunc} is the unique c-function with these two properties. Eq.~\eqref{toyc} agrees with the central charge obtained in \cite{Li:2018kqp}, $\ell(1-\g\chi_s^2/\ell^2)$ in the normalization used there, upon identifying $\chi_s^2 = -(1+\L_0\ell^2)/(2\a)$. Remarkably, the same combination is the coefficient of the kinetic term of the linearized graviton modes around this vacuum \cite{Li:2018rgn}, so that for this model positivity of the central charge coincides with the absence of ghosts. We will come back to this in Sections~\ref{sec:posc} and \ref{sec:discussion}.

As an independent check, we now compute the central charge from the logarithmic divergence of the on-shell action, by defining the boundary theory on S$^2$. For this purpose, we consider the Euclidean metric
\begin{equation}\label{toyeuc}
	\dd{s}^2 = \frac{\dd{r}^2}{1+\frac{r^2}{\ell^2}} + r^2 \left(\dd{\th}^2 + \sin^2 \th \dd{\vf}^2\right),
\end{equation}
which is the Euclidean continuation of AdS$_3$, i.e.\ the hyperbolic space H$^3$ of radius $\ell$, with the boundary S$^2$ at $r\to\infty$. First we need to find the scalar configuration supporting this solution. We therefore take the more general ansatz
\begin{equation}\label{eucansgen}
	\dd{s}^2 = \cN^2(r) \dd{r}^2 + \cR^2(r) \left(\dd{\th}^2 + \sin^2 \th \dd{\vf}^2\right),
\end{equation}
which is the most general metric with $SO(3)$ symmetry, and assume $\f=\f(r)$. The remarks made above for the domain wall apply here as well: $\cN$ plays the role of the lapse and its equation is the Hamiltonian constraint, so both functions must be kept in the reduced action for $\Phi_A = \{\cN, \cR, \f\}$ and varied. After deriving the Euler-Lagrange equations, we use the reparametrization freedom to set $\cR = r$, and look for a solution with $\cN = \frac{1}{\sqrt{1+r^2/\ell^2}}$, which is the metric \eqref{toyeuc}. The equations are satisfied provided the scalar field is
\begin{equation}\label{toyphiE}
	\f = \sqrt{-\frac{1+\L_0 \ell^2}{2\a}}\, \log\left[\frac{r^2 + \ell^2}{r_0^2}\right], \qquad r_0: \text{constant},
\end{equation}
where we have used the condition $\g = \a \ell^2$. Note that on this metric the equation of motion of $\f$ is identically satisfied, since $G^{\m\n} = \frac{1}{\ell^2} g^{\m\n}$ on H$^3$ and hence $\a g^{\m\n} - \g G^{\m\n} = 0$ for $\g = \a \ell^2$. The profile of $\f$ is fixed by the constraint equation following from the variation with respect to $\cN$.

The central charge is proportional to the logarithmic divergence of the Euclidean on-shell action $I_E = - \int \dd^3 x \sqrt{g}\, \mathcal{L}$,\footnote{The minus sign comes from the Wick rotation $t = -i\tau$ of the Lorentzian action $I = \int \dd^3 x \sqrt{-g}\,\mathcal{L}$, $iI = -I_E$, so that the partition function is $Z = e^{-I_E}$. With this sign the coefficient of the logarithm is negative, as required for $\log Z$ on S$^2$ to grow as $+\frac{c}{3}\log(r_c/\ell)$ for a theory with central charge $c$.} evaluated with a UV cut-off at $r = r_c$. The other divergences are expected to be cancelled by surface terms and appropriate counterterms \cite{Balasubramanian:1999re,Emparan:1999pm}, and we do not need them. For the same reason as above, the scalar terms in \eqref{actg} vanish identically on the solution, and the on-shell Lagrangian is simply
\begin{equation}\label{toyLon}
	\mathcal{L} = R - 2\L_0 = -\frac{2}{\ell^2}\left(3 + \L_0 \ell^2\right).
\end{equation}
Since this is a constant, the logarithm comes only from the volume element. The angular integral gives the area of the unit sphere, $\int \dd\th\, \dd\vf \sin\th = 4\pi$, and
\begin{equation}\label{toyvol}
	\int \dd^3 x \sqrt{g} = 4\pi \int_0^{r_c} \dd{r}\, \frac{r^2}{\sqrt{1 + r^2/\ell^2}}
	= 2\pi \ell\, r_c^2 - 2\pi \ell^3 \log\left(\frac{r_c}{\ell}\right) + \dots ,
\end{equation}
where the dots denote finite and vanishing terms. Therefore,
\begin{equation}\label{toyIlog}
	\Ilog = -4\pi \ell \left(3 + \L_0 \ell^2\right) \log\left(\frac{r_c}{\ell}\right).
\end{equation}
For GR ($\L_0 \ell^2 = -1$) the same computation gives $\Ilog = -8\pi \ell \log\left(r_c/\ell\right)$, corresponding to the central charge \eqref{cGR}. Since the logarithmic divergence is linear in the central charge, the ratio of the two coefficients gives
\begin{equation}\label{toyratio}
	\frac{c}{c_{\rm GR}} = \frac{3 + \L_0 \ell^2}{2}\,,
\end{equation}
which reproduces exactly the central charge \eqref{toyc} obtained from the c-function.

\section{General proof}\label{sec:proof}
\subsection{Vacuum solution and the c-function}\label{sec:vacuum}
The regularization of the Lovelock invariants \eqref{Ln} to a Horndeski Lagrangian was carried out at arbitrary order in \cite{Colleaux:2020wfv}. Evaluated in 3d, and in the normalization of \cite{Fernandes:2025eoc}, which we adopt, the $n$-th order regularized Lovelock invariant reads
\begin{equation}\label{regLov}
	\cL^{(n)} =  G_2^{(n)}(\phi, X)-G_3^{(n)}(\phi, X) \square \phi+G_4^{(n)}(\phi, X) R +G_{4 X}^{(n)}\left[(\square \phi)^2-\left(\nabla_\mu \nabla_\nu \phi\right)^2\right],
\end{equation}
where
\begin{equation}
	G_2^{(n)}=-2^{n+1}(n-1) X^n, \qquad G_3^{(n)}=2^n n X^{n-1}, \qquad G_4^{(n)}=-\frac{2^{n-1} n}{2 n-3} X^{n-1},
\end{equation}
and $X=- \frac{1}{2}\partial_\mu \phi \partial^\mu \phi$. We consider the full tower of these invariants, coupled to the same matter sector as before,
\begin{equation}\label{actLov}
	I = \int \dd^3 x \sqrt{-g} \Bigg[ R - 2 \L_{0} + \frac{1}{L^2} \sum_{n=2}^{\infty} c_n L^{2n} \cL^{(n)} - \frac{1}{2}  (\pd \chi)^2 \Bigg],
\end{equation}
where $L$ is a fixed length scale and the $c_n$ are dimensionless coupling constants. The AdS radius $\ell$ of the vacuum is determined by $\L_0$ and the $c_n$, and is in general different from $L$.

We proceed exactly as in Section~\ref{sec:toy}: we insert the domain wall ansatz \eqref{domgen} with $\f = \f(r)$ and $\chi = \chi(r)$ into the action, derive the Euler-Lagrange equations of the reduced action for $\Phi_A = \{A, B, \f, \chi\}$, and then gauge fix $B = 0$. Note that $X = -\f'^2/2 < 0$ for a radial profile, so that $X^k = (-1)^k \f'^{2k}/2^k$. This is the origin of the factors $(-1)^n$ below. The equations of motion in the gauge $B=0$ read
\begin{align}
	0 &= 4\L_0 + \chi'^2 + 4 A'^2 + 4 A'' \nn \\
	& \quad - 4 \sum_{n=2}^{\infty} (-1)^n c_n L^{2n-2} \f'^{2n-3} \Big[ n \f' \left(A'^2 + A''\right) - (n-1) \f'^3 + 2n(n-1) \left(A' - \f'\right) \f'' \Big], \label{AeqnL}\\
	0 &= 4\L_0 - \chi'^2 + 4 A'^2 - 4 \sum_{n=2}^{\infty} (-1)^n c_n L^{2n-2} \f'^{2n-2} \Big[ n(2n-1) A'^2 - 4n(n-1) A' \f' + (n-1)(2n-1) \f'^2 \Big], \label{BeqnL}\\
	0 &= \frac{\dd}{\dd r} \Bigg[ e^{2A} \sum_{n=2}^{\infty} n(n-1)(-1)^n c_n L^{2n-2} \f'^{2n-3} \left(A' - \f'\right)^2 \Bigg], \label{phieqnL}\\
	0 &= 2 \chi' A' + \chi''. \label{chieqnL}
\end{align}
For $n=2$ these reduce to the equations of the 3d Gauss-Bonnet model \cite{Hennigar:2020fkv}. As in Section~\ref{sec:toy}, the equation for $\f$ is a total derivative and integrates to
\begin{equation}\label{firstintL}
	e^{2A} \sum_{n=2}^{\infty} n(n-1)(-1)^n c_n L^{2n-2} \f'^{2n-3} \left(A' - \f'\right)^2 = \l, \qquad \l = \text{constant}.
\end{equation}
The remarkable feature of \eqref{firstintL} is that the factor $(A' - \f')^2$ is common to all orders $n$. As in Section~\ref{sec:toy}, $\l$ is the conserved charge of the shift symmetry of $\f$, and $J_\m J^\m \propto \l^2 e^{-4A}$ diverges at the Poincar\'e horizon of the vacuum unless $\l = 0$. We restrict again to this sector, as in \cite{Alkac:2022zda}. Eq.~\eqref{firstintL} then has two branches. On the branch $\f' = 0$ all the Lovelock terms vanish and one is back to GR, with $\L_0 = -1/\ell^2$ for any couplings. On the other branch, since the sum in \eqref{firstintL} does not vanish for generic couplings, we conclude that
\begin{equation}\label{phiA}
	\f' = A',
\end{equation}
along the entire flow, for any choice of the $c_n$. In contrast to the model of Section~\ref{sec:toy}, the warp factor $A$ is no longer frozen to its vacuum value. Instead, the Horndeski scalar is locked to it. Substituting \eqref{phiA} into \eqref{AeqnL} and \eqref{BeqnL} gives
\begin{align}
	0 &= 4\L_0 + \chi'^2 + 4 A'^2 + 4 A'' - 4 \sum_{n=2}^{\infty} (-1)^n c_n L^{2n-2} \left( A'^{2n} + n A'^{2n-2} A'' \right), \label{AeqnLA}\\
	0 &= 4\L_0 - \chi'^2 + 4 A'^2 - 4 \sum_{n=2}^{\infty} (-1)^n c_n L^{2n-2} A'^{2n}. \label{BeqnLA}
\end{align}

The vacuum AdS$_3$ solution is obtained by setting $\chi = 0$ and $A = r/\ell$, so that $\f = r/\ell$ up to a constant. Eq.~\eqref{BeqnLA} gives the relation between the bare cosmological constant, the AdS radius and the couplings,
\begin{equation}\label{vacL}
	1 + \L_0 \ell^2  = \sum_{n=2}^{\infty} (-1)^n \hat{c}_n, \qquad \hat{c}_n \equiv c_n \left( \frac{L}{\ell} \right)^{2n-2},
\end{equation}
and \eqref{AeqnLA} is then automatically satisfied. The dimensionless combinations $\hat{c}_n$ measure the couplings in units of the AdS radius and will appear in all the results below. In the limit $c_n \to 0$, one recovers $\L_0 = -1/\ell^2$. Note also that the scalar profile $\f = r/\ell$ is linear in $r$, as in Section~\ref{sec:toy}. It therefore breaks the special conformal transformations of AdS$_3$, and since the tower depends on $\f$ only through $X$, this holds at every order and for any values of the $c_n$: the dual theory is scale-invariant but not conformal.

To study the NEC, we eliminate $\L_0$ between Eqs.~\eqref{AeqnLA} and \eqref{BeqnLA}, which gives
\begin{equation}\label{FL}
	F = - A'' \left[ 1 - \sum_{n=2}^{\infty} n(-1)^n \hat{c}_n \left( \ell A' \right)^{2n-2} \right] \geq 0.
\end{equation}
For $c_n = 0$ this is the GR result $-A'' \geq 0$. Following the same logic as in Section~\ref{sec:toy}, we look for a c-function whose derivative is related to $F$ as in GR, $c'(r) = F/(\ell_p A'^2)$, and which reduces to $1/(\ell_p A')$ when the couplings are switched off. The unique c-function that meets both requirements is
\begin{equation}\label{cfuncL}
	c(r) = \frac{1}{\ell_p A'} \left[ 1 + \sum_{n=2}^{\infty} \frac{n(-1)^n}{2n-3}\, \hat{c}_n \left( \ell A' \right)^{2n-2} \right],
\end{equation}
since, using $\frac{\dd}{\dd r} A'^{2n-3} = (2n-3) A'^{2n-4} A''$,
\begin{equation}\label{cprimeL}
	c'(r) = -\frac{A''}{\ell_p A'^2} \left[ 1 - \sum_{n=2}^{\infty} n(-1)^n \hat{c}_n \left( \ell A' \right)^{2n-2} \right] = \frac{F}{\ell_p A'^2}  \geq 0.
\end{equation}
It is unique because $c'(r)$ is fixed by the first requirement and the GR limit fixes the integration constant. The c-function \eqref{cfuncL} is therefore monotonically increasing towards the UV for any values of the couplings, as a direct consequence of the NEC. The factors $1/(2n-3)$ in \eqref{cfuncL} come from the integration in $A'$. They are absent from $F$.

Evaluating \eqref{cfuncL} at the UV fixed point $ A' = 1/\ell$, we obtain the central charge
\begin{equation}\label{cinfL}
	c_\infty = \frac{\ell}{\ell_p} \left[ 1 + \sum_{n=2}^{\infty} \frac{n(-1)^n}{2n-3}\, \hat{c}_n \right].
\end{equation}
For $n=2$ this reads $c_\infty = \frac{\ell}{\ell_p}\left(1 + 2\hat{c}_2\right)$, and for $c_n = 0$ one recovers \eqref{cGR}. Note that \eqref{cinfL} is linear in the couplings, as is the relation \eqref{vacL} between $\L_0$ and $\ell$. The c-function itself, however, depends on the couplings non-linearly through $A'(r)$ along the flow. In the next subsection we verify \eqref{cinfL} by computing the central charge from the on-shell action.

\subsection{Central charge from the on-shell action}\label{sec:anomaly}
As in Section~\ref{sec:toy}, we compute the logarithmic divergence of the Euclidean on-shell action $I_E = -\int \dd^3 x \sqrt{g}\,\mathcal{L}$ for the theory defined on S$^2$, using the ansatz \eqref{eucansgen} with $\f = \f(r)$ and a UV cut-off at $r=r_c$. Inserting \eqref{eucansgen} into the action \eqref{actLov}, deriving the Euler-Lagrange equations for $\Phi_A = \{\cN, \cR, \f\}$ and gauge fixing $\cR = r$ afterwards, we find that the equation for $\cN$ is algebraic,
\begin{equation}\label{ENL}
	0 = \L_0 \cN^2 r^2 - \cN^2 + 1 + \sum_{n=2}^{\infty} (-1)^n c_n L^{2n-2} \left( \frac{\f'}{\cN} \right)^{2n-2} \Big[ n \cN^2 - (2n-1)(n-1) r^2 \f'^2 + 4n(n-1) r \f' - n(2n-1) \Big],
\end{equation}
and that the equation for $\f$ is again a total derivative, which integrates to
\begin{equation}\label{JL}
	\sum_{n=2}^{\infty} n(n-1)(-1)^n c_n L^{2n-2} \left( \frac{\f'}{\cN} \right)^{2n-2} \frac{1}{\cN \f'}\left[ \frac{\cN^2}{2n-3} - \left(r \f' - 1\right)^2 \right] = \text{constant}.
\end{equation}
Smoothness of the solution at the origin requires $\cN(0) = 1$, so that there is no conical defect, and $\f'(0) = 0$, since an $SO(3)$-symmetric scalar that is smooth at the origin must be an even function of $r$. Every term on the left-hand side of \eqref{JL} is then proportional to $\f'^{2n-3}$ at $r=0$ and vanishes, so the constant is zero. The equation for $\cR$ is implied by these two, by the same reparametrization argument given for $B$ in Section~\ref{sec:toy}.

\paragraph{The $n=2$ case.} For $n=2$, the system can be solved exactly. Eq.~\eqref{JL} has two branches, $\f' = 0$, which is the GR vacuum, and $\cN^2 = (r\f' - 1)^2$. On the latter, the root regular at the origin is $\cN = 1 - r \f'$. Substituting this into \eqref{ENL} and using $\L_0 = -\frac{1}{\ell^2}(1 - \hat{c}_2)$ from \eqref{vacL}, one obtains a quadratic equation for $\cN^2$ whose root that is continuously connected to GR is
\begin{equation}\label{n2exact}
	\cN = \frac{\ell}{\sqrt{r^2 + \ell^2}}, \qquad \f' = \frac{1 - \cN}{r}, \qquad \f = \log\left(\ell + \sqrt{r^2 + \ell^2}\right) + \text{constant}.
\end{equation}
The metric is therefore exactly H$^3$, as in the model of Section~\ref{sec:toy}, but now with a scalar profile that has $\f' = 1/r + \mathcal{O}(1/r^2)$, matching the linear profile $\f = A$ of the Poincar\'e vacuum. In contrast to Section~\ref{sec:toy}, the scalar does not drop out of the on-shell Lagrangian, which is now a non-trivial function of $r$. Inserting \eqref{n2exact} into the reduced Lagrangian and expanding at large $r$, we find\footnote{The exact expression is lengthy and not illuminating. Up to a total derivative, $\cL^{(2)}$ equals $-\left[4G^{\m\n}\pd_\m\f\pd_\n\f - 4(\pd\f)^2\square\f + 2(\pd\f)^4\right]$, the 3d Gauss-Bonnet invariant of \cite{Hennigar:2020fkv,Alkac:2022fuc} with $\a = -c_2 L^2$. Using this form one finds the compact expression $\cN r^2 \mathcal{L} = \frac{4\left[\hat{c}_2 \ell^3 - \hat{c}_2 \ell^2\sqrt{r^2+\ell^2} - \ell r^2\right]}{\ell^2 \sqrt{r^2+\ell^2}}$. The two forms differ in the power-law terms but have the same $1/r$ coefficient, since a total derivative cannot produce a logarithm when the fields have a Laurent expansion in $r$.}
\begin{equation}\label{n2integrand}
	\cN r^2 \mathcal{L} = -4\left(1 + 4 \hat{c}_2\right) \frac{r}{\ell} + 12 \hat{c}_2 + 2\left(1 + 2\hat{c}_2\right) \frac{\ell}{r} + \mathcal{O}\!\left(\frac{1}{r^2}\right).
\end{equation}
The first two terms are power-law divergences. The $1/r$ term gives the logarithm,
\begin{equation}\label{n2Ilog}
	\Ilog = -8\pi \ell \left(1 + 2\hat{c}_2\right) \log\left(\frac{r_c}{\ell}\right),
\end{equation}
where the factor $4\pi$ comes from the angular integral and the minus sign from the Euclidean action. Comparing with the GR result $\Ilog = -8\pi\ell\log(r_c/\ell)$ of Section~\ref{sec:toy},
\begin{equation}\label{n2ratio}
	\frac{c}{c_{\rm GR}} = 1 + 2\hat{c}_2,
\end{equation}
which is exactly \eqref{cinfL} truncated to $n=2$. Note that here, unlike in Section~\ref{sec:toy}, the logarithm does not come from the volume of $H^3$ alone. Writing $\mathcal{L} = \mathcal{L}_0 + \mathcal{L}_1/r + \mathcal{L}_2/r^2 + \dots$ and $\cN r^2 = \ell r - \ell^3/(2r) + \dots$, the coefficient of $1/r$ in the integrand is $-\frac{\ell^3}{2}\mathcal{L}_0 + \ell \mathcal{L}_2$. In Section~\ref{sec:toy} the on-shell Lagrangian was constant and only the first term contributed. Here $\mathcal{L}_0 = -\frac{4}{\ell^2}(1+4\hat{c}_2)$, and the first term gives $2\ell(1+4\hat{c}_2)$, which is not the central charge. It is the second term, $\ell\mathcal{L}_2 = -4\hat{c}_2\ell$, coming from the $1/r^2$ tail of the on-shell Lagrangian, that corrects it to $2\ell(1+2\hat{c}_2)$. Evaluating the boundary value of the on-shell Lagrangian on the regularized volume of $H^3$ would therefore give the wrong answer.

\paragraph{Asymptotic expansion.} For $n \geq 3$, or for several orders switched on at once, the metric supporting the S$^2$-symmetric solution is no longer exactly H$^3$: the scalar backreacts, and one finds $\cN^{-2} = r^2/\ell^2 + 1 + \mathcal{O}(1/r)$ with a non-vanishing $1/r$ term. A closed form is not available. It is, however, not needed. The logarithmic divergence of $I_E = -4\pi\int_0^{r_c} \dd r\, \cN r^2 \mathcal{L}$ is produced by the $1/r$ term of the radial integrand alone: writing $\cN r^2 \mathcal{L} = a\, r + b + \frac{e}{r} + \mathcal{O}(1/r^2)$ at large $r$, one has $\int_0^{r_c} \dd r\, \cN r^2 \mathcal{L} = \frac{a}{2} r_c^2 + b\, r_c + e \log r_c + \text{finite}$, where the finite part contains all the information about the interior of the solution. Therefore it is enough to know the solution to a fixed order at large $r$. We write
\begin{equation}\label{asympL}
	\cN = \frac{\ell}{r}\left[ 1 + \nu_1 \frac{\ell}{r} + \nu_2 \frac{\ell^2}{r^2} + \dots \right], \qquad
	r \f' = 1 + p_1 \frac{\ell}{r} + p_2 \frac{\ell^2}{r^2} + \dots,
\end{equation}
where the leading terms are fixed by the AdS asymptotics with radius $\ell$ and, on the branch with a non-trivial scalar, by the requirement that \eqref{JL} be finite (the bracket must be of order $1/r^2$, which forces $r\f' \to 1$). The coefficients are determined order by order from \eqref{ENL} and \eqref{JL}.

As a warm-up, we redo the $n=2$ case in this way. Expanding \eqref{ENL} in powers of $\ell/r$, the first three orders give
\begin{equation}\label{n2orders}
	1 + \ell^2 \L_0 - \hat{c}_2 = 0, \qquad \nu_1 \left( 1 - 2\hat{c}_2 \right) = 0, \qquad \left( 2\nu_2 + 1 \right)\left( 1 - 2\hat{c}_2 \right) = 0,
\end{equation}
i.e.\ the vacuum relation, $\nu_1 = 0$ and $\nu_2 = -1/2$. Eq.~\eqref{JL} at leading order gives $p_1^2 = 1$, and at the next order $p_2 = \frac{1}{2}(1 - p_1^2) = 0$. The branch $p_1 = -1$ reproduces the expansion of the exact solution \eqref{n2exact}. Inserting these into the reduced Lagrangian one recovers \eqref{n2integrand}, and hence \eqref{n2Ilog}. In fact, the $1/r$ coefficient of $\cN r^2 \mathcal{L}$ does not depend on $p_1$ and $p_2$ at all. Only $\nu_1 = 0$ and the vacuum relation are needed.

\paragraph{The full tower.} For the full theory we proceed in the same way. Expanding \eqref{ENL} to the first three orders in $\ell/r$ gives
\begin{align}
	0 &= 1 +  \L_0  \ell^2- \sum_{n=2}^{\infty} (-1)^n \hat{c}_n, \label{ord0}\\
	0 &= \nu_1 \Big[ 1 - \sum_{n=2}^{\infty} n(-1)^n \hat{c}_n \Big], \label{ord1}\\
	0 &= \left(2\nu_2 + 1\right) \Big[ 1 - \sum_{n=2}^{\infty} n(-1)^n \hat{c}_n \Big], \label{ord2}
\end{align}
where the coefficients $p_1, p_2$ of the scalar do not appear. The first equation is the vacuum relation \eqref{vacL}, and the other two give $\nu_1 = 0$ and $\nu_2 = -1/2$ for arbitrary couplings, as long as the common factor, which is the same effective coupling that multiplies $A''$ in \eqref{FL} at $\ell A'=1$, is non-zero. The leading order of \eqref{JL} fixes $p_1^2$ as a coupling-weighted average,
\begin{equation}\label{p1L}
	p_1^2 \sum_{n=2}^{\infty} n(n-1)(-1)^n \hat{c}_n = \sum_{n=2}^{\infty} \frac{n(n-1)(-1)^n}{2n-3}\, \hat{c}_n,
\end{equation}
and the higher orders determine $p_2$, $\nu_3$, etc.\ as non-linear functions of the couplings. For a single order $n$, for example, $\nu_3 \propto (n-2)$, which is why H$^3$ is exact only for $n=2$.

Inserting \eqref{asympL} into the reduced Lagrangian and expanding, we find for the coefficient of $1/r$
\begin{equation}\label{masterL}
	\left[\cN r^2 \mathcal{L}\right]_{1/r} = 2\ell \left\{ 1 + \sum_{n=2}^{\infty} \frac{n(-1)^n}{2n-3}\,\hat{c}_n + \nu_1^2 \Big[ 1 - \sum_{n=2}^{\infty} n(-1)^n \hat{c}_n \Big] - \nu_2 \Big[ 1 + \ell^2\L_0 - \sum_{n=2}^{\infty} (-1)^n \hat{c}_n \Big] \right\},
\end{equation}
where we have kept $\nu_1$, $\nu_2$ and $\L_0$ free. Remarkably, this coefficient does not depend on $p_1$ and $p_2$: each of the four terms in $\cL^{(n)}$ contributes such terms, but they cancel in the sum. Moreover, the coefficient of $\nu_2$ is precisely the vacuum relation \eqref{ord0}. Using \eqref{ord0} and \eqref{ord1}, the $1/r$ coefficient collapses to $2\ell\left[1 + \sum_n \frac{n(-1)^n}{2n-3} \hat{c}_n\right]$, so that
\begin{equation}\label{IlogL}
	\Ilog = -8\pi \ell \left[ 1 + \sum_{n=2}^{\infty} \frac{n(-1)^n}{2n-3}\, \hat{c}_n \right] \log\left(\frac{r_c}{\ell}\right).
\end{equation}
Comparing again with the GR result,
\begin{equation}\label{ratioL}
	\frac{c}{c_{\rm GR}} = 1 + \sum_{n=2}^{\infty} \frac{n(-1)^n}{2n-3}\, \hat{c}_n,
\end{equation}
which is exactly the value $c_\infty$ of the c-function \eqref{cinfL} at the UV fixed point, for arbitrary values of the couplings $c_n$. This completes the proof of the holographic c-theorem for the full tower of regularized Lovelock invariants.

Let us stress what was actually needed for this result: the first two orders \eqref{ord0}, \eqref{ord1} of the constraint and the expansion \eqref{masterL} of the on-shell Lagrangian. Every quantity that could carry information about the interior of the solution, or that depends non-linearly on the couplings ($p_1$, $p_2$, $\nu_3$, \dots), drops out of the logarithm. This is the concrete form of the statement that the central charge is fixed by the asymptotic data alone: the AdS radius, the curvature of the boundary S$^2$, the unit coefficient of the $\log r$ behaviour of the scalar, and the couplings.

\section{Resummation}\label{sec:resum}
The results of the previous section hold for arbitrary values of the couplings $c_n$. Particular choices of the $c_n$ have received a lot of attention recently, because the infinite sums can be performed in closed form. The idea originated in 4d quasi-topological gravity, where the resummed theories admit regular black holes \cite{Bueno:2024dgm}. It was then carried over to the regularized Lovelock tower. In 4d, the choices of couplings were introduced in \cite{Fernandes:2025fnz}, where the cosmological singularity was shown to be resolved, and regular black holes with a well-defined thermodynamics were obtained in \cite{Cisterna:2025vxk}. In 3d, the resummed theories were studied in \cite{Fernandes:2025eoc}, where regular versions of the BTZ black hole and their thermodynamics were obtained. In this section we study the choices of couplings of \cite{Fernandes:2025eoc}, together with a one-parameter deformation of one of them, and give the corresponding resummed Lagrangians, c-functions and central charges.

\subsection{Generating functions}\label{sec:gen}
All the quantities computed in Section~\ref{sec:proof} depend on the couplings only through the combinations $(-1)^n c_n$, multiplied by powers of $L A'$ or $L/\ell$. It is therefore convenient to introduce the generating function
\begin{equation}\label{hdef}
	h(x) \equiv \sum_{n=2}^{\infty} (-1)^n c_n\, x^n ,
\end{equation}
together with
\begin{equation}\label{gdef}
	g(x) \equiv \sum_{n=2}^{\infty} \frac{n(-1)^n}{2n-3}\, c_n\, x^{n-1} = \frac{\sqrt{x}}{2} \int_0^x \dd t\, \frac{h'(t)}{t^{3/2}},
\end{equation}
where the integral representation follows from $\int_0^x \dd t\, t^{n-5/2} = 2x^{n-3/2}/(2n-3)$. In terms of these, the vacuum relation \eqref{vacL}, the NEC quantity \eqref{FL}, the c-function \eqref{cfuncL} and the central charge \eqref{cinfL} read
\begin{align}
	1 + \ell^2 \L_0 &= \frac{h(y)}{y}, \qquad y \equiv \frac{L^2}{\ell^2}, \label{vach}\\
	F &= -A''\, P(x), \qquad P(x) \equiv 1 - h'(x), \qquad x \equiv L^2 A'^2, \label{Fh}\\
	c(r) &= \frac{1}{\ell_p A'} \left[ 1 + g(x) \right], \label{ch}\\
	c_\infty &= \frac{\ell}{\ell_p} \left[ 1 + g(y) \right]. \label{cinfh}
\end{align}

The same two functions give the resummed Lagrangian. Collecting the Einstein-Hilbert term, the cosmological constant and the tower in \eqref{actLov} into a single Horndeski Lagrangian of the form \eqref{regLov}, and using $X^k = (-1)^k u^k/(2L^2)^k$ with
\begin{equation}\label{udef}
	u \equiv -2L^2 X = L^2\, \pd_\m \f \pd^\m \f ,
\end{equation}
one finds
\begin{equation}\label{Gresum}
	G_2 = -2\L_0 + \frac{2}{L^2}\left[ h(u) - u\, h'(u) \right], \qquad G_3 = -2 h'(u), \qquad G_4 = 1 + g(u).
\end{equation}
On the domain wall $u = L^2 A'^2 = x$, so that \eqref{ch} can be written as
\begin{equation}\label{cG4}
	c(r) = \frac{G_4}{\ell_p A'}\bigg|_{\text{flow}} .
\end{equation}
This holds order by order and does not require the resummation: the c-function is the GR c-function multiplied by the function $G_4$ that multiplies the Ricci scalar in the Lagrangian, i.e.\ by the effective Planck mass, evaluated on the flow. A similar statement holds for the pure-metric theories of \cite{Chernicoff:2024lkj}, whose central charge is the inverse of an effective Planck length. Likewise, $P = 1 + \frac12 G_3$ on the flow.

We note that, on the flow, $P = G_4 - 2X\pd_X G_4$ order by order, so that the c-function and its monotonicity are controlled by two different combinations of $G_4$ and its $X$-derivative. The c-function \eqref{cG4} also has a geometric reading. A covariant c-function built from the extrinsic curvature $K$ of codimension-two slices of the bulk was recently proposed in \cite{Jokela:2026cjm}. In three bulk dimensions it reduces to $1/(G_3 K)$ up to a numerical factor, and on the domain wall \eqref{dom} the slice of constant $t$ and $r$ has $K = A'$, so that this is the GR c-function \eqref{cGRfunc}. In the form \eqref{regLov} of the Lagrangian, the Riemann tensor appears only through $G_4 R$, so the Wald entropy density \cite{Wald:1993nt} of the same slice, $\frac{\pd \cL}{\pd R_{\m\n\r\s}}\epsilon_{\m\n}\epsilon_{\r\s}$ with $\epsilon_{\m\n}$ the binormal in the $t$-$r$ plane, is $G_4$ in units of its GR value. Eq.~\eqref{cG4} is then the geometric c-function with $1/G_3$ replaced by the Wald density of the slice, which is the prescription used to construct c-functions in higher-curvature gravity \cite{Sinha:2010ai,Myers:2010tj,Paulos:2010ke}. This identification depends on the form of the Lagrangian, as observed in \cite{Alkac:2022zda}. Forms that differ by a total derivative give the same field equations and the same c-function, but not the same $\pd\cL/\pd R_{\m\n\r\s}$.

\subsection{Positivity of the central charge and the structure of the vacua}\label{sec:posc}
In this section, by vacuum we mean the AdS$_3$ vacuum with the linear scalar profile of Section~\ref{sec:vacuum}. The GR vacuum $\f' = 0$ with $\L_0 = -1/\ell^2$ exists in every case and is not affected by the couplings. Two functions of the couplings therefore control the physics, and each has a clear meaning. The first is $1 + g$, the effective Planck mass. By \eqref{cG4}, $c_\infty > 0$ is the requirement that $G_4$ be positive on the vacuum,
\begin{equation}\label{posc}
	1 + g(y) > 0 .
\end{equation}
This is the analogue of requiring a positive Newton constant in GR, since $G_4$ on the vacuum is the inverse of the effective Newton constant in units of $G_3$: $c_\infty = \ell\, G_4|_{\rm vac}/\ell_p$. For $G_4 < 0$ the gravitational sector has the wrong overall sign on the vacuum, and we discard such vacua. For the model of Section~\ref{sec:toy}, this condition is exactly the no-ghost condition for the linearized modes found in \cite{Li:2018rgn}. Positivity of the central charge is also what one would require of the dual theory if it were a CFT, where $c > 0$ is necessary for unitarity, since the norm of the state $L_{-2}|0\rangle$ is $c/2$. As discussed in Section~\ref{sec:toy}, the dual theory here is only scale-invariant, so we do not rely on this argument and regard \eqref{posc} as a condition on the bulk vacuum. It translates into a condition on the couplings and on $\L_0$, which we determine below for each choice of couplings.

The second function is the effective coupling $P$, which already appeared in Section~\ref{sec:proof}: it multiplies $A''$ in the NEC quantity, and its value at the UV fixed point is the common factor in Eqs.~\eqref{ord1}, \eqref{ord2}. It also controls the vacuum. Writing the vacuum relation as $\L_0 L^2 = h(y) - y$, one finds
\begin{equation}\label{dLam}
	\frac{\dd (\L_0 L^2)}{\dd y} = - P(y),
\end{equation}
so that the bare cosmological constant is a monotonic function of the AdS radius exactly when $P$ does not change sign. In that case the vacuum is unique: for a given $\L_0$ there is one AdS radius $\ell$. When $P$ changes sign, there are several AdS vacua for the same $\L_0$, in analogy with the Einstein and non-Einstein branches of Gauss-Bonnet gravity.\footnote{In Einstein-Gauss-Bonnet gravity in $d \geq 5$, the equation for the effective cosmological constant of the AdS vacuum is quadratic and has two roots. One of them reduces to the GR value when the Gauss-Bonnet coupling is sent to zero (the Einstein branch). The other diverges in this limit, and the graviton propagating around it has a kinetic term of the wrong sign \cite{Boulware:1985wk}. Eq.~\eqref{dLam} shows that $P$ plays the same role here: it changes sign exactly where two branches of vacua meet.} Since $F = \frac{1}{2}\chi'^2 \geq 0$, the sign of $P$ along the flow also fixes the sign of $A''$: for $P > 0$ one has $A'' \leq 0$ as in GR, while for $P < 0$ the NEC forces $A'' \geq 0$. In both cases the c-function is monotonic, $c' = \chi'^2/(2\ell_p A'^2) \geq 0$. The two functions are related by $P = 1 + g - 2x\, g'$, which is just the statement $c' = F/(\ell_p A'^2)$, but the two conditions $1 + g > 0$ and $P > 0$ are independent, and we will see examples where one holds and the other fails.

The infinite sums \eqref{hdef}, \eqref{gdef} have a finite radius of convergence, and the resummed expressions below define the theory beyond it by analytic continuation. This is where the interesting physics lies: the singularities of $h$ on the positive real axis are genuine boundaries of the theory, while singularities on the negative real axis are never reached for real $A'$.

\subsection{Examples}\label{sec:examples}
We now consider the four choices of couplings listed in Table I of \cite{Fernandes:2025eoc}, and a deformation of the first one. In all of them $c_1 = 1$, so that the tower includes the Einstein-Hilbert term as its $n=1$ member, consistently with $G_4^{(1)} = 1$ in \eqref{regLov}. Our sums start at $n=2$. For each choice we give $h$, $P$ and $g$, the vacuum relation, the central charge, and the condition for its positivity. The resummed functions $G_2$, $G_3$, $G_4$ obtained from \eqref{Gresum} are collected in Table~\ref{tab:G}, and the results in Table~\ref{tab:results}.

\paragraph{$c_n = 1$.} The generating function is geometric,
\begin{equation}
	h(x) = \frac{x^2}{1+x}, \qquad P(x) = \frac{1}{(1+x)^2}, \qquad g(x) = \frac{x}{2(1+x)} + \frac{3}{2}\sqrt{x}\,\arctan\sqrt{x},
\end{equation}
and the vacuum relation takes a remarkably simple form,
\begin{equation}\label{case1vac}
	\L_0 = -\frac{1}{\ell^2 + L^2}.
\end{equation}
An AdS vacuum therefore exists only if $|\L_0| < 1/L^2$, with $\ell^2 = 1/|\L_0| - L^2$, and it is unique since $P > 0$. The c-function and the central charge are
\begin{equation}\label{case1}
	c(r) = \frac{1}{\ell_p A'}\left[ 1 + \frac{L^2 A'^2}{2\left(1 + L^2 A'^2\right)} + \frac{3}{2} L A' \arctan\left(L A'\right) \right], \qquad
	c_\infty = \frac{\ell}{\ell_p}\left[ 1 + \frac{L^2}{2\left(\ell^2 + L^2\right)} + \frac{3L}{2\ell} \arctan\frac{L}{\ell} \right].
\end{equation}
Although the series converge only for $L A' < 1$, the resummed expressions are smooth for all $A' > 0$: the only singularity of $h$ is at $x = -1$. Both $1+g$ and $P$ are positive for all $x$, so the central charge is positive for any value of the couplings. The higher-order terms raise it, $c_\infty > \ell/\ell_p$, and they also bound it from below: as $|\L_0| \to 1/L^2$ the AdS radius shrinks to zero, but $c_\infty \to 3\pi L/(4\ell_p)$ stays finite. The central charge can thus be set by the scale $L$ of the tower rather than by the AdS radius.

\paragraph{$c_n = q^{-n}$.} This one-parameter deformation of the previous case, which we introduce here, interpolates between different behaviours as $q$ is varied. One has $h(x) = x^2/[q(q+x)]$ and
\begin{equation}
	P(x) = 1 - \frac{1}{q} + \frac{q}{(q+x)^2}, \qquad g(x) = \frac{1}{q}\left[ \frac{v}{2(1+v)} + \frac{3}{2}\sqrt{v}\,\arctan\sqrt{v} \right], \qquad v \equiv \frac{x}{q},
\end{equation}
with the vacuum relation
\begin{equation}\label{case5vac}
	\L_0 L^2 = -\frac{y\left[ q^2 + (q-1) y \right]}{q\,(q+y)}.
\end{equation}
Since $g > 0$, the central charge is positive for all $q$ and all couplings. The parameter $q$ instead controls the vacuum structure. For $q \geq 1$ the effective coupling is positive, $\L_0 L^2$ is monotonic and the vacuum is unique. For $q>1$ all negative $\L_0$ are allowed, while for $q = 1$ one recovers the bound $|\L_0| < 1/L^2$. For $q < 1$, $P$ becomes negative for large $x$: $\L_0 L^2$ decreases from zero to a minimum and then grows to $+\infty$, so that for $\L_0$ between the minimum and zero there are two AdS vacua, one continuously connected to GR and one with $P<0$, while for $\L_0 \geq 0$ there is a single AdS vacuum with $P<0$, i.e.\ a negative effective cosmological constant is generated from a vanishing or positive bare one. On the $P<0$ branch the central charge is positive and the c-function is monotonic, but the NEC forces $A'' \geq 0$, the opposite of GR. In analogy with Gauss-Bonnet gravity, we regard the branch connected to GR as the physical one.

\paragraph{$c_n = 1/n$.} Here
\begin{equation}
	h(x) = x - \log(1+x), \qquad P(x) = \frac{1}{1+x}, \qquad g(x) = \sqrt{x}\,\arctan\sqrt{x},
\end{equation}
and again the vacuum relation can be solved in closed form,
\begin{equation}\label{case2vac}
	\L_0 L^2 = -\log\left(1 + \frac{L^2}{\ell^2}\right) \qquad \Longleftrightarrow \qquad \ell^2 = \frac{L^2}{e^{|\L_0| L^2} - 1}.
\end{equation}
In contrast to the first case, every negative value of $\L_0$ gives an AdS vacuum, unique since $P>0$. The c-function and the central charge are
\begin{equation}\label{case2}
	c(r) = \frac{1}{\ell_p A'}\left[ 1 + L A' \arctan\left(L A'\right) \right], \qquad
	c_\infty = \frac{\ell}{\ell_p}\left[ 1 + \frac{L}{\ell}\arctan\frac{L}{\ell} \right],
\end{equation}
positive for all couplings and smooth for all $A'$. The tower again raises the central charge and bounds it from below: $c_\infty \to \pi L/(2\ell_p)$ as $|\L_0| \to \infty$, where $\ell \to 0$.

\paragraph{$c_n = \frac{1 - (-1)^n}{2n}$.} Only odd $n$ contribute, with $c_n = 1/n$, and since $(-1)^n = -1$ for these terms the alternating sign in \eqref{hdef} is lost:
\begin{equation}
	h(x) = x - \operatorname{arctanh} x, \qquad P(x) = \frac{1}{1 - x^2}, \qquad g(x) = -\frac{\sqrt{x}}{2}\left[ \operatorname{arctanh}\sqrt{x} - \arctan\sqrt{x} \right].
\end{equation}
The function $h$ is singular at $x = 1$, on the positive real axis, so the theory only makes sense for $L A' < 1$. The vacuum relation is again solvable,
\begin{equation}\label{case3vac}
	\L_0 L^2 = -\operatorname{arctanh}\frac{L^2}{\ell^2} \qquad \Longleftrightarrow \qquad \ell^2 = \frac{L^2}{\tanh\left(|\L_0| L^2\right)},
\end{equation}
which shows that the AdS radius is automatically larger than $L$ for every $\L_0 < 0$, and that $\ell \to L$ as $|\L_0| \to \infty$: the AdS radius is bounded from below by the scale of the tower. Since $P > 0$ on the whole domain, the vacuum is unique. The higher-order terms now lower the central charge, $g<0$, and its positivity becomes an issue: $1 + g$ vanishes at $y_* \simeq 0.985$, so that $c_\infty > 0$ requires $y < y_*$, which by \eqref{case3vac} means
\begin{equation}
	|\L_0| L^2 < \operatorname{arctanh} y_* \simeq 2.45 .
\end{equation}
For larger values of $|\L_0|$ the effective Planck mass on the vacuum is negative, and the vacuum is discarded.

\paragraph{$c_n = \frac{\Gamma(n/2)\,\delta_{0,k}}{\sqrt{\pi}\,\Gamma\left(\frac{n+1}{2}\right)}$.} Here $k = (n-1) \bmod 2$ and $\delta$ is the Kronecker delta, so that again only odd $n$ contribute. The couplings are the Taylor coefficients of $(1-z^2)^{-1/2}$, $c_{2m+1} = \binom{2m}{m}/4^m$, and
\begin{equation}
	h(x) = x - \frac{x}{\sqrt{1-x^2}}, \qquad P(x) = \frac{1}{(1-x^2)^{3/2}}, \qquad 1 + g(x) = {}_2F_1\!\left(-\tfrac{1}{4}, \tfrac{3}{2}; \tfrac{3}{4}; x^2\right),
\end{equation}
where the hypergeometric function is the resummation of the series \eqref{gdef}. This case is qualitatively the same as the previous one: the alternating sign is lost, $h$ is singular at $x=1$, $P>0$ on the whole domain and the vacuum is unique. The vacuum relation is
\begin{equation}\label{case4vac}
	\L_0 L^2 = -\frac{y}{\sqrt{1-y^2}} \qquad \Longleftrightarrow \qquad \ell^4 = L^4 + \frac{1}{\L_0^2},
\end{equation}
so that again $\ell > L$ for every $\L_0<0$, with $\ell \to L$ as $|\L_0| \to \infty$. The tower lowers the central charge, and $1+g$ vanishes at $y_* \simeq 0.906$, and positivity requires
\begin{equation}
	|\L_0| L^2 < \frac{y_*}{\sqrt{1-y_*^2}} \simeq 2.14 .
\end{equation}

\begin{table}[t]
	\centering
	\renewcommand{\arraystretch}{1.9}
	\begin{tabular}{c c c c}
		\toprule
		$c_n$ & $\dfrac{L^2}{2}\left(G_2 + 2\L_0\right)$ & $G_3$ & $G_4$ \\
		\midrule
		$1$ & $-\dfrac{u^2}{(1+u)^2}$ & $-\dfrac{2u(u+2)}{(1+u)^2}$ & $1 + \dfrac{u}{2(1+u)} + \dfrac{3}{2}\sqrt{u}\arctan\sqrt{u}$ \\
		$q^{-n}$ & $-\dfrac{u^2}{(q+u)^2}$ & $-\dfrac{2u(u+2q)}{q(q+u)^2}$ & $1 + \dfrac{1}{q}\left[\dfrac{v}{2(1+v)} + \dfrac{3}{2}\sqrt{v}\arctan\sqrt{v}\right],\ v = \dfrac{u}{q}$ \\
		$\dfrac{1}{n}$ & $\dfrac{u}{1+u} - \log(1+u)$ & $-\dfrac{2u}{1+u}$ & $1 + \sqrt{u}\arctan\sqrt{u}$ \\
		$\dfrac{1-(-1)^n}{2n}$ & $\dfrac{u}{1-u^2} - \operatorname{arctanh} u$ & $\dfrac{2u^2}{1-u^2}$ & $1 - \dfrac{\sqrt{u}}{2}\left[\operatorname{arctanh}\sqrt{u} - \arctan\sqrt{u}\right]$ \\
		$\dfrac{\Gamma(n/2)\,\delta_{0,k}}{\sqrt{\pi}\,\Gamma(\frac{n+1}{2})}$ & $\dfrac{u^3}{(1-u^2)^{3/2}}$ & $\dfrac{2}{(1-u^2)^{3/2}} - 2$ & ${}_2F_1\!\left(-\tfrac{1}{4}, \tfrac{3}{2}; \tfrac{3}{4}; u^2\right)$ \\
		\bottomrule
	\end{tabular}
	\caption{Resummed Horndeski functions of the theory \eqref{actLov} for the five choices of couplings, from Eq.~\eqref{Gresum}, with $u = L^2 \pd_\m\f\pd^\m\f$ and $k = (n-1) \bmod 2$. On the domain wall $u = L^2 A'^2$, and $G_4$ becomes the factor $1+g$ of the c-function.}
	\label{tab:G}
\end{table}

\begin{table}[t]
	\centering
	\renewcommand{\arraystretch}{1.9}
	\resizebox{\textwidth}{!}{%
		\begin{tabular}{c c c c c}
			\toprule
			$c_n$ & $P(x)$ & vacuum relation & AdS vacuum & $c_\infty > 0$ \\
			\midrule
			$1$ & $\dfrac{1}{(1+x)^2}$ & $\L_0 = -\dfrac{1}{\ell^2+L^2}$ & unique, $|\L_0| < \dfrac{1}{L^2}$ & always \\
			$q^{-n}$ & $1 - \dfrac{1}{q} + \dfrac{q}{(q+x)^2}$ & $\L_0 L^2 = -\dfrac{y\left[q^2 + (q-1)y\right]}{q(q+y)}$ & unique ($q \geq 1$), several ($q<1$) & always \\
			$\dfrac{1}{n}$ & $\dfrac{1}{1+x}$ & $\ell^2 = \dfrac{L^2}{e^{|\L_0|L^2} - 1}$ & unique, any $\L_0<0$ & always \\
			$\dfrac{1-(-1)^n}{2n}$ & $\dfrac{1}{1-x^2}$ & $\ell^2 = \dfrac{L^2}{\tanh(|\L_0|L^2)}$ & unique, any $\L_0<0$, $\ell > L$ & $|\L_0| L^2 < 2.45$ \\
			$\dfrac{\Gamma(n/2)\,\delta_{0,k}}{\sqrt{\pi}\,\Gamma(\frac{n+1}{2})}$ & $\dfrac{1}{(1-x^2)^{3/2}}$ & $\ell^4 = L^4 + \dfrac{1}{\L_0^2}$ & unique, any $\L_0<0$, $\ell > L$ & $|\L_0| L^2 < 2.14$ \\
			\bottomrule
		\end{tabular}%
	}
	\caption{Summary of the results for the five choices of couplings: the effective coupling $P$, the relation between the bare cosmological constant and the AdS radius, the structure of the AdS vacua, and the condition for the positivity of the central charge $c_\infty = \frac{\ell}{\ell_p}\left[1+g(y)\right]$, with $x = L^2A'^2$, $y = L^2/\ell^2$ and $k = (n-1) \bmod 2$.}
	\label{tab:results}
\end{table}

\subsection{Comments}\label{sec:comments}
The five examples illustrate the general structure. Since all the results of Section~\ref{sec:proof} depend on the couplings only through $h$ and $g$, any choice of $c_n$ whose generating function can be resummed gives a closed-form Lagrangian, vacuum relation, c-function and central charge, and the matching between $c_\infty$ and the coefficient of the logarithmic divergence holds automatically, because it holds order by order and both sides are linear in the $c_n$. Whether the resummed theory is well behaved is controlled by three properties: the location of the singularities of $h$, which must lie on the negative real axis for the couplings to be unrestricted; the positivity of $1 + g$ on the vacuum, which is the positivity of the effective Planck mass and of the central charge; and the positivity of $P = 1 - h'$, which makes the vacuum unique and the geometry GR-like ($A'' \leq 0$). The choices in which all orders contribute with positive couplings ($c_n = 1$, $c_n = 1/n$, $c_n = q^{-n}$ with $q \geq 1$) satisfy all three properties for any value of the couplings. In these theories the higher-order terms raise the central charge, and bound it from below by a number of order $L/\ell_p$ set by the scale of the tower rather than by the AdS radius. The two choices with odd orders only, in which the alternating sign in \eqref{hdef} is lost, have a singularity of $h$ at $x=1$, a unique vacuum with $\ell > L$, and a central charge that is lowered by the tower and becomes negative beyond a critical value of $|\L_0| L^2$ of order two. Finally, the deformation $c_n = q^{-n}$ with $q<1$ is the only case with several vacua and a non-GR-like branch.

\section{Discussion and outlook}\label{sec:discussion}
We have shown that 3d Lovelock gravity admits a holographic c-theorem at all orders in the tower of regularized Lovelock invariants, for arbitrary values of the couplings. The proof relies on two facts that hold order by order in the minisuperspace of domain wall solutions: the equation of the Horndeski scalar is a total derivative, which forces $\f' = A'$ along the flow, and the NEC then takes the form $F = -A''P$, which can be integrated once in $A'$. The UV value of the c-function is linear in the couplings and matches the coefficient of the logarithmic divergence of the on-shell action on S$^2$. Since the scalar profile of the vacuum breaks the special conformal transformations, the dual theory is scale-invariant rather than conformal. All the results depend on the couplings only through two generating functions, which allowed us to resum them for the choices of couplings of \cite{Fernandes:2025eoc}. The function $G_4$ multiplying the Ricci scalar in the resummed Lagrangian is the c-function up to the GR factor $1/(\ell_pA')$, so positivity of the central charge is positivity of the effective Newton constant on the vacuum.

Two questions deserve further study. The first is unitarity. In the bulk, it should be studied by expanding the action to second order in the fluctuations of the metric and of the scalar around the vacuum $(g_{\rm AdS}, \f = r/\ell)$, and checking the signs of the kinetic terms of the propagating modes. This would show whether bulk unitarity and positivity of the central charge are compatible, unlike in NMG \cite{Bergshoeff:2009aq,Liu:2009bk}. On the boundary, the question is more subtle because the dual theory is not conformal. In 2d, a unitary theory that is Poincar\'e and scale-invariant, and has a discrete spectrum, is necessarily conformal \cite{Polchinski:1987dy, Nakayama:2013is}. The dual of 3d Lovelock gravity, being scale-invariant but not conformal, must therefore either be non-unitary or violate one of these assumptions. The latter is plausible: the shift symmetry of $\f$ acts on the boundary as a non-compact symmetry, which typically leads to a continuous spectrum. In a scale-invariant theory that is not conformal the trace of the stress tensor is the divergence of a virial current, which here might be related to the current of the shift symmetry. Both points can be tested holographically, through the Ward identity for dilatations and the two-point functions of the stress tensor and of the operator dual to $\f$. For the model of Section~\ref{sec:toy} these two-point functions were computed in \cite{Li:2018rgn}: the trace mode decouples in $d=2$, no local virial current was identified, and the coefficient of $\langle TT\rangle$ was found to agree with the central charge only to linear order in the coupling, a discrepancy that remains to be understood. Extending this analysis to the full tower is the obvious continuation.

The second concerns the regular black holes of the resummed theories \cite{Fernandes:2025eoc}. For hairy black holes in 3d, the ground state is not global AdS$_3$ but the soliton obtained by a double Wick rotation, and the entropy is reproduced by a version of the Cardy formula that makes no reference to the central charge \cite{Correa:2010hf,Correa:2011dt}. In \cite{Alkac:2024hvu}, we showed that it works for the static black holes of 3d Lovelock gravity at the orders where the solutions were known, and in \cite{Cisterna:2025vxk} it was shown to work for the 4d regular black holes with an infinite tower of corrections when the horizon is planar. This has now been extended to all orders in 3d, including the regular black holes of the resummed theories and the rotating case \cite{Ozsahin:2026eao}: the Cardy formula without central charges reproduces the entropy for arbitrary couplings, and the effective central charge defined by the soliton mass, $c_\text{eff}=\frac{3\ell}{2G_3}\left[1 - \frac{h(y)}{y}\right]^{-1}$, differs from the central charge $c_\infty$ of Section~\ref{sec:resum}. The relation between the two, and whether $c_\infty$ can be read off from the black hole spectrum in some other way, remains to be understood.

\paragraph{Acknowledgements}
We thank Aimeric Coll\'eaux, Mario Flory and Niko Jokela for correspondence and comments on the first version of this paper. G.A. thanks Mustafa \c{C}\"o\c{c}elli for suggesting the use of Claude for the challenging parts of this project. L.G. is partially supported by the Agencia Nacional de Investigaci\'on y Desarrollo (ANID) through Fondecyt Iniciaci\'on grant No.~11260910. During this work, the authors used the large language model Claude Fable 5.1 (Anthropic) to verify the field equations, first integrals and asymptotic expansions by symbolic computation, to carry out parts of the derivations under the authors' direction, to check the numerical values quoted in Section~\ref{sec:resum}, and to draft and edit parts of the manuscript. All results were independently checked by the authors, who take full responsibility for the content of the paper.

\bibliographystyle{utphys}
\bibliography{ref}

\end{document}